\documentclass[epj]{svjour}
\usepackage{graphics}
\begin{document}
\journalname{Eur. Phys. J. A}

\title{Alpha-decay spectroscopy of $^{257}$Rf}

\author{
K. Hauschild\inst{1}\thanks{\emph{e-mail:} k.hauschild@ijclab.in2p3.fr} \and
A. Lopez-Martens\inst{1} \and 
R. Chakma\inst{1}\thanks{\emph{Present address:} GANIL, 14076 Caen Cedex 5, France} \and
O. Dorvaux\inst{2} \and 
B.J.P. Gall\inst{2} \and 
M.L. Chelnokov\inst{3} \and 
V.I. Chepigin\inst{3} \and 
A.V. Isaev\inst{3} \and 
I.N. Izosimov\inst{3} \and 
D.E. Katrasev\inst{3} \and 
A.A. Kuznetsova\inst{3} \and 
O.N. Malyshev\inst{3} \and 
A.G. Popeko\inst{3} \and 
Yu.A. Popov\inst{3} \and 
E.A. Sokol\inst{3} \and 
A.I. Svirikhin\inst{3} \and 
M.S. Tezekbayeva\inst{3}\inst{4} \and 
A.V. Yeremin\inst{3} \and 
D. Ackermann\inst{5} \and 
J. Piot\inst{5} \and 
P. Mosat\inst{6} \and 
B. Andel\inst{6}
}
\institute{JCLab, IN2P3-CNRS, Universit\'e Paris Saclay, F-91405 Orsay, France \label{addr1}
           \and
           Universit\'e de Strasbourg, CNRS, IPHC UMR 7178, F-67000 Strasbourg \label{addr2}
           \and
           FLNR, JINR, Dubna, Russia\label{addr3}
		  \and
           The Institute of Nuclear Physics, 050032, Almaty, The Republic of Kazakhstan\label{addr4}
		  \and
           GANIL, 14076 Caen Cedex 5, France\label{addr5}
		  \and
           Department of Nuclear Physics and Biophysics, Comenius University in Bratislava, 84248 Bratislava, Slovakia\label{addr6}
}
\date{Received: date / Revised version: date}
%
\abstract{
The decay properties of states in $^{257}$Rf have been investigated with the detector array GABRIELA at the FLNR, Dubna. The electromagnetic decay of a new excited state in $^{253}$No has been observed. The state lies 750 keV above the ground state and is favourably populated in the alpha decay from a low-lying isomeric state in $^{257}$Rf. It decays to the 9/2$^-$ ground state by an M1 transition and is assigned the 11/2$^-$[725] Nilsson configuration. The presence of this state suggests a possible reinterpretation of the decay of the high-K isomer in $^{253}$No. Due to the favoured nature of the $\alpha$-decay the 11/2$^-$[725] Nilsson configuration is also assigned to the first excited state of $^{257}$Rf, lying at 74 keV.
} 
\maketitle
\section{Introduction}
\label{intro}
The spectrum of excited states in super heavy nuclei is the result of a delicate balance between the strong Coulomb repulsion between the numerous protons in the nucleus and the properties of the strong force acting between the many nucleons in the system. Following specific states across isotonic and isotopic states chains can highlight how the system rearranges itself as neutrons or protons are added and can reveal effects such as changes in the underlying single-particle spectra and coupling with collective phenomena. 

Combined prompt and decay spectroscopy by means of alpha, gamma and internal conversion measurements has proven to be a powerful tool to investigate decay patterns of excited nuclear states in transfermium nuclei (see reference \cite{Ackermann} for a full review). In this paper, the decay properties of the ground state and an isomeric state of $^{257}$Rf are investigated using the Si and Ge detector array GABRIELA \cite{Hauschild} installed at the focal plane of the Separator for Heavy ELement Spectrosopy (SHELS) \cite{Popeko}. 

\section{Experimental details}

To produce $^{257}$Rf nuclei, the reaction $^{208}$Pb($^{50}$Ti,1n) was used at cyclotron beam energies of 240 to 245 MeV and  Pb target thicknesses of 400-450 $\mu$g/cm$^{2}$. The targets had a 1.5 $\mu$m Ti backing and were rotated to withstand the intense flux of $^{50}$Ti ions provided by the U400 cyclotron at the FLNR, JINR. While a total dose of 4.68$\times$10$^{18}$ incident particles was recorded during the experiment the beam energy was not always optimised for the production of $^{257}$Rf since the 2n evaporation channel $^{256}$Rf was also of interest.

\begin{figure*}[hbtp]
   \centering
   \resizebox{0.95\textwidth}{!}{%
      \includegraphics{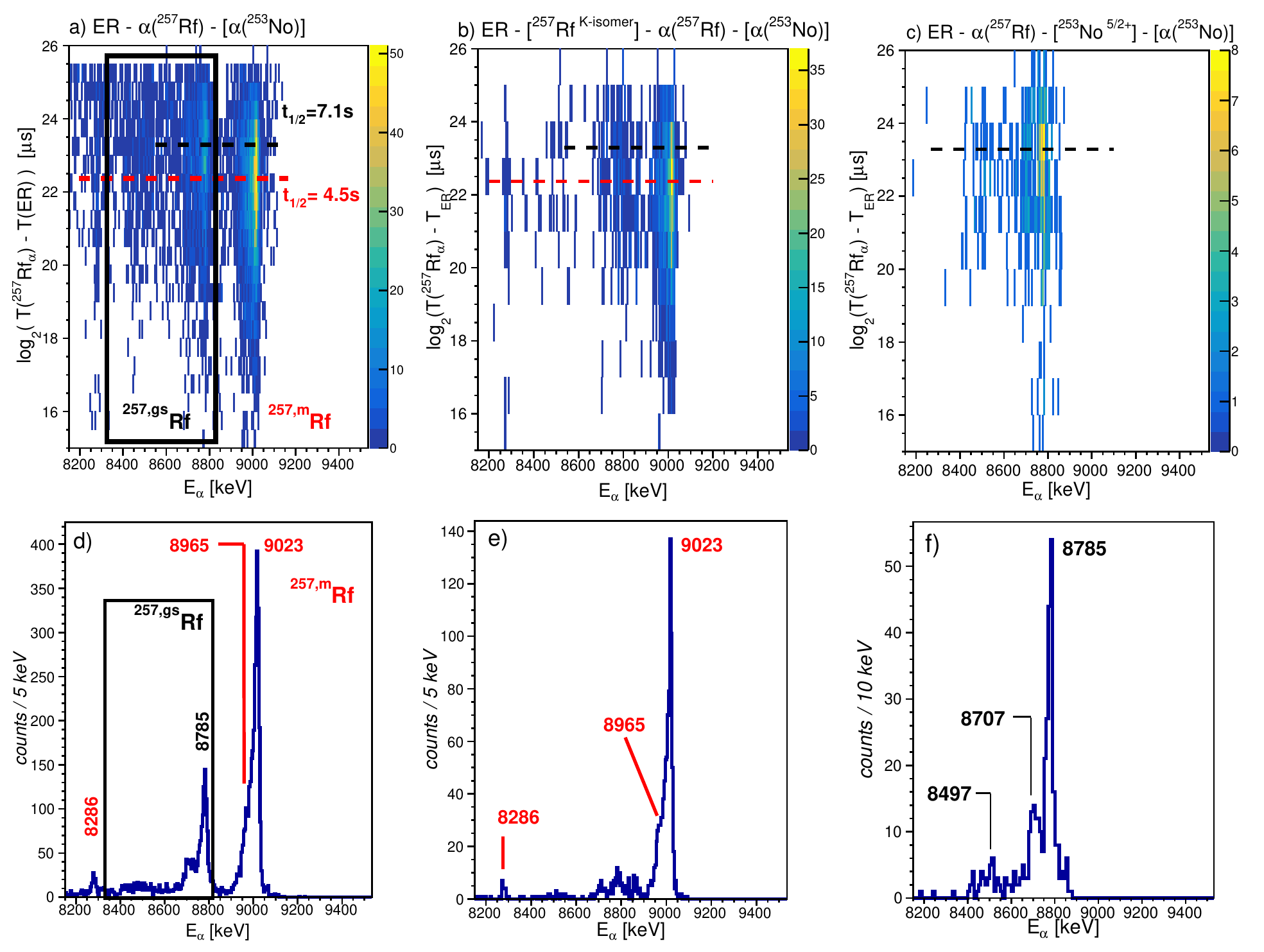}
    }
   \caption{[Colour online] The time between the implantation of an evaporation residue (ER) and the detection of the first alpha decay in the same pixel as a function of the detected alpha-particle energy. In addition, a) was followed by the characteristic alpha decay of $^{253}$No, b) was preceeded by the decay of the K-isomer in $^{257}$Rf \cite{Rissanen} and then followed by the characteristic alpha decay of $^{253}$No, and c) was followed by both the deexcitation of the 5/2$^+$ isomer in $^{253}$No and then the characteristic alpha decay of $^{253}$No. Panels d), e) and f) are the energy projections of a), b) and c), respectively. The area within the black rectangle in a) and d) is dominated by decays from the ground state with its apparent half-life marked by a dashed black line. The transitions labelled in red are from the excited meta-stable state in $^{257}$Rf with the dashed red line indicating the observed half-life.}
   \label{fig1}
\end{figure*}

The nuclei to be studied were separated by SHELS and implanted into the 10x10 cm$^2$ Double Sided Silicon-strip Detector (DSSD) of the newly upgraded GABRIELA detector array \cite{Chakma}. This detector provides 16384 different pixels for genetic position and time correlations between implantation and subsequent decay events. Surrounding the DSSD, in the upstream direction, 8 smaller-sized DSSDs forming a tunnel were used to detect escaping $\alpha$ particles and fission fragments as well as internal conversion electrons (ICEs).  Gamma and X rays were detected with a ring of 4 coaxial Ge detectors and a large clover detector placed just behind the DSSD. The signals from the individual crystals of the clover can be added together, the so called``add-back" mode \cite{Duchene}, to increase the high-energy gamma-ray efficiency. The Ge detectors were surrounded by a dedicated BGO shield, which provided a flag for coincident Ge and BGO events, and allowed for Compton scattered events to be removed in software in order to improve the signal to background ratio.
The signals from all the detectors were time-stamped with a 1$~\mu$s clock. The amplification range of the analogue electronics was set to detect ICEs and $\alpha$ particles up to 25 MeV on both the front and back side of the DSSD.

Energy and efficiency calibration runs for conversion electrons and $\alpha$ particles were performed with $^{50}$Ti-induced reactions on $^{164}$Dy and  $^{150}$Er targets leading to the production of $^{209m,210m}$Ra \cite{Ressler,Hauschild08} and the $\alpha$-emitting isotopes $^{216}$Th \cite{Hauschild01}, $^{212}$Ra and $^{208}$Rn \cite{Rytz}. A correction of 7 keV has been applied to the $\alpha$ particle energies of $^{257}$Rf  to account for the different recoil energy contribution to the measured total energy of the calibration nuclei compared to $^{257}$Rf: see \cite{Huang} for more details. The Ge detectors were calibrated with standard $^{133}$Ba and $^{152}$Eu sources.

\section{Results}

\begin{figure}[htbp]
   \centering
      \resizebox{0.95\columnwidth}{!}{%
		\includegraphics{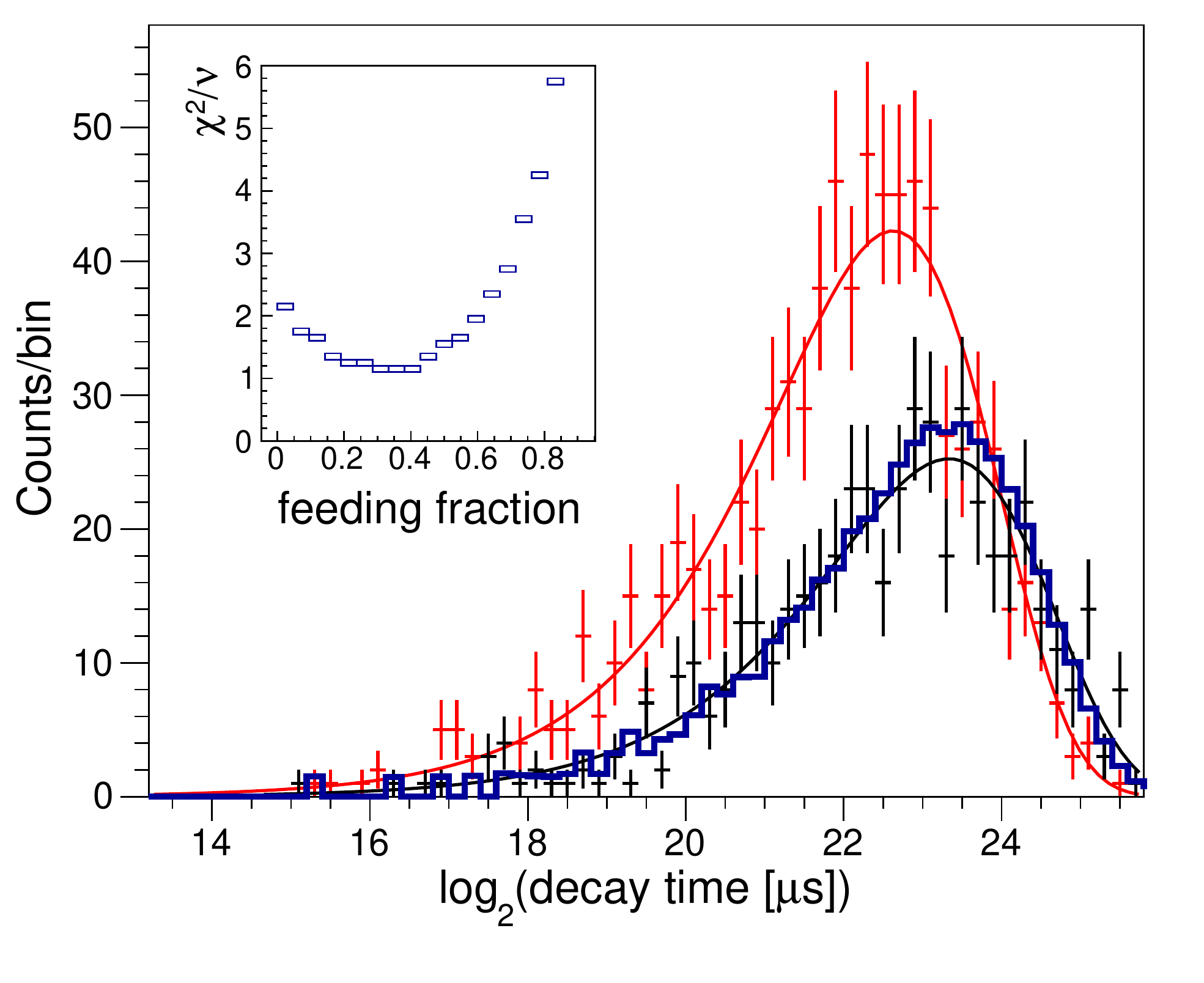}
	}
   \caption{[Colour online] Logarithmic decay time distributions of the alpha-particle groups shown in Fig. \ref{fig1}: black for the ground state, $^{257,gs}$Rf, and red for the excited meta-stable state, $^{257,m}$Rf. The solid lines are the fitted single component exponential decay curves. The inset shows the $\chi^2$ statistic as a function of the fractional feeding of the ground state via the $^{257,m}$Rf state.
The solid blue line is the result of the simulation with the minimum $\chi^2$ having a factional feeding of 0.35. See text for details.}
   \label{log2T}
\end{figure}

The time between the implantation of an evaporation residue (ER) and the detection of the $^{257}$Rf alpha decay in the same pixel as a function of the detected alpha-particle energy is shown in Fig. \ref{fig1} (a) and the projection on to the energy axis in Fig. \ref{fig1} (d). To ensure these spectra represent clean $^{257}$Rf events the additional condition that a position correlated $^{253}$No alpha decay was observed to follow the $^{257}$Rf decay was made. Two groups of lines with clearly different lifetimes are visible; one between 8.3 and 8.85 MeV and the other below 8.3 MeV and between 8.9 and 9.1 MeV.  They correspond to the emission from the ground state (gs) and a low-lying metastable state ($^{257,m}$Rf) respectively \cite{Hessberger,Qian,Streicher,Berryman,Rissanen}. The corresponding time distributions (data points) and single component fits (solid curves) \cite{Lopez} are shown in Fig. \ref{log2T} : black for the ground-state and red for the excited state. The apparent half-lives are also given in Table \ref{tab1}.
The half-life of $^{257,m}$Rf agrees well with the literature values. The half-life of the ground state, however, is found to be 1.6 times larger than that in the ENSDF evaluation \cite{ENSDF} but is consistent with the measurement of Dragojevic et al., \cite{Dragojevic}. Simulations were carried out to check the possibility that the ground state has an apparent lifetime due to feeding from the $^{257,m}$Rf state via internal decay. The true half-life of the ground state was taken to be the value measured when the ground state is populated via the alpha decay of $^{261}$Sg. The most recently measured value is 5.6 s \cite{Streicher}. The apparent life-time is then simulated for varying percentages of feeding from the higher lying isomeric state (in steps of 5\%) and a decay curve is constructed for each assumed feeding percentage. A $\chi^2$ statistic is then calculated between the measured and simulated decay curves, presented in insert of Fig. \ref{log2T} as a function of feeding percentage, which shows a distinct minimum at 35(5)\% feeding. The corresponding decay curve in Fig. \ref{log2T} of this ``best fit" is the sold blue line which visually follows the experimental data (black data points) more closely than the fit using a single decay component (black curve).

\begin{table}[tbh]
\caption{(Apparent) Half-lives obtained in this work for different states in $^{257}$Rf and associated $\alpha$ decays.}

\begin{tabular}{c l l r r}
\hline 
\hline\noalign{\smallskip}
Nuclear state &  $t_{1/2}$ [s] ~ & ~E$_\alpha$ [keV]~ & ~I$_{\alpha}$ [\%]~ & hindrance\\
\noalign{\smallskip}\hline\noalign{\smallskip}

$^{257,gs}$Rf & 7.1(3)  & 8785(5)    & *\\
                       &            & 8497(15) & * \\
$^{257,m}$Rf  & 4.5(1) & 9023(5)   &  7.1(18) & 241(69)\\
                       &            & 8965(5)  &  89($^{21}_{23}$) & 13(4)\\
                       &             & 8286(5)  & 3.9(12) & 2.1(7)\\

\hline
\hline
\end{tabular}
\label{tab1}
\\
$*$ reliable values could not be obtained due to summing effects
\end{table}

It was found that requesting the detection of the K-isomer reported by Rissanen et al. \cite{Rissanen} in the same DSSD pixel prior to the $^{257}$Rf $\alpha$-decay enhanced the alpha decay from the metastable state over that from the ground state. This can be seen by comparing Figs \ref{fig1} (a,d) with (b,e). In particular, the line at 8286 in Fig \ref{fig1} (d) survives this condition. From Fig. \ref{fig1}(e)  an estimate of the intensity ratio of the two $\alpha$ groups decaying from the low-lying isomer was extracted as N$_\alpha$(8286):N$_\alpha$(8965+9023) = 23(5):885(30).

The alpha decay from the ground state of $^{257}$Rf is correlated to the internal decay of the 5/2$^+$ isomer \cite{Bemis,Lopez} at 167 keV above the ground state of $^{253}$No. A prominent peak in this group is found at 8785 keV, as previously reported in references  \cite{Hessberger,Streicher}. This is illustrated clearly in Figs. \ref{fig1} c and f) which present the  alpha-particle energy spectrum for $^{257}$Rf  conditioned by recoil $-$ $\alpha$($^{257}$Rf) $-$ $^{253}$No-isomer $-$  $\alpha$($^{253}$No) correlations within the same DSSD pixel. Other peaks that are enhanced by this additional constraint are observed at 8497 and 8707 keV. However, the complexity of the decay feeding into the 5/2$^+$[622] level, resulting in summing effects, does not allow reliable branching ratios for the alpha decay from the ground state of $^{257}$Rf to be obtained.

\begin{figure*}[htbp]
   \centering
   \resizebox{0.95\textwidth}{!}{%
      \includegraphics{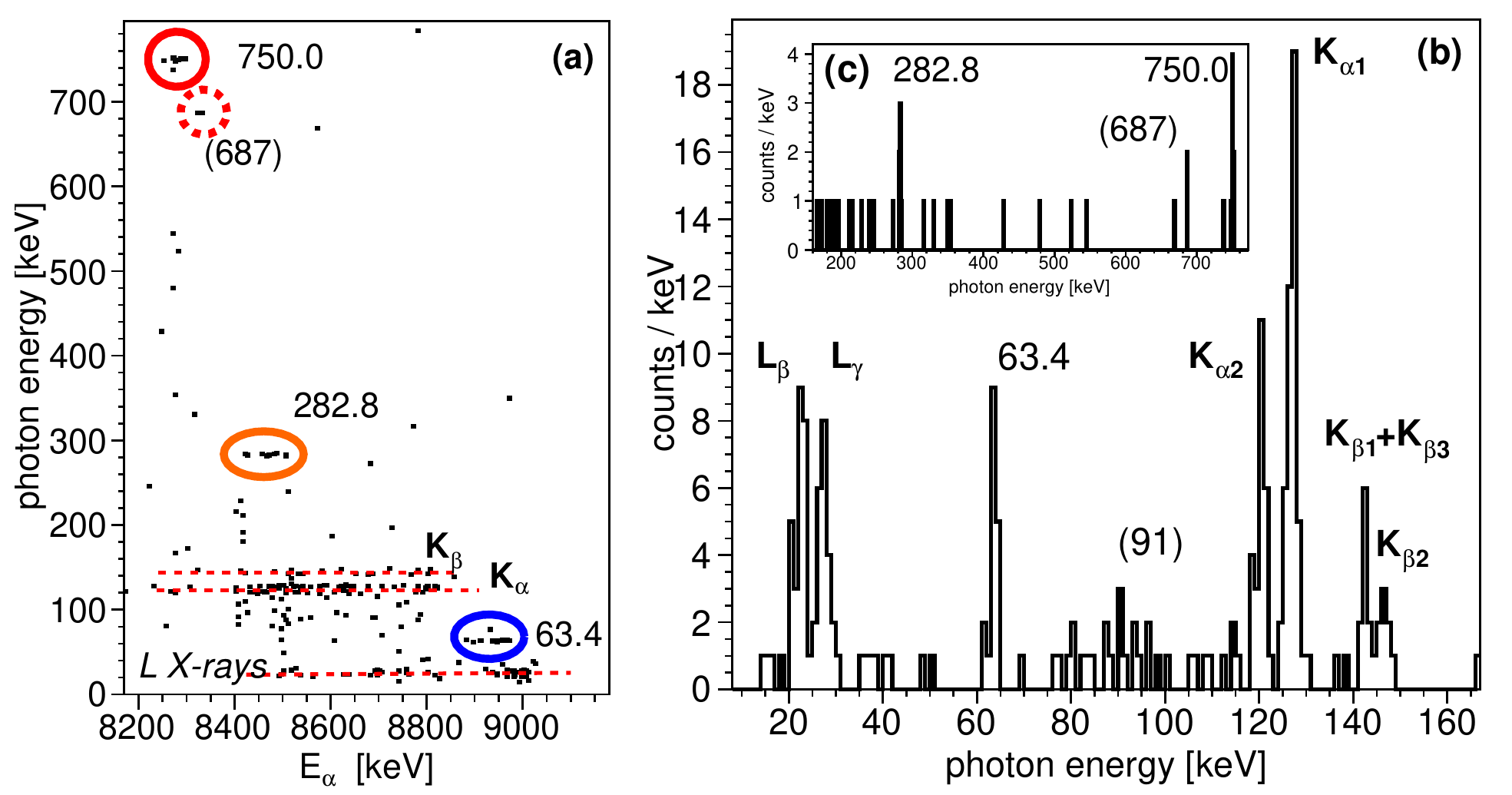}
    }
   \caption{[Colour online] a) Compton-suppressed and $\gamma$-ray energy``add-back'' (see text for details)  spectrum observed in prompt coincidence with correlated $^{257}$Rf alpha particles with the additional constraint that full energy $^{253}$No were subsequently position correlated: ER - [$\alpha$($^{257}$Rf)-$\gamma$] - $\alpha$($^{253}$No). b) Low energy and (c) high energy portions of the projected photon energy. Transition energies are labelled in keV.}
   \label{alpha_gamma}
\end{figure*}

Fig. \ref{alpha_gamma} shows the Compton-suppressed $\gamma$ ray spectrum observed in prompt coincidence with first generation alpha particles which were subsequently correlated to the characteristic alpha decay of $^{253}$No between 7.95 and 8.13 MeV: i.e. ER - [$\alpha$($^{257}$Rf)-$\gamma$] - $\alpha$($^{253}$No). The three lines observed at 750, 283 and 63 keV are therefore assigned to the decay of excited states in $^{253}$No. No evidence for the 557-keV transition reported by Hessberger et al., \cite{Hessberger2} is observed. There is some evidence of the 91-keV line observed by Streicher et al., \cite{Streicher} in Fig. \ref{alpha_gamma} (b), however it is coincident with a broad alpha particle energy distribution similar to that of the observed K X-rays and is therfore inconsistent with the level scheme proposed by Streicher. 
The 63-keV transition is fed by the 8965-keV alpha decay from $^{257,m}$Rf (15 counts) and agrees with the energy reported for the 11/2$\rightarrow$9/2  transition in the ground state rotational band \cite{Herzberg,Mistry}. The 750-keV transition (10 counts) has not been observed following $^{257}$Rf $\alpha$-decay before. The state emitting the 750-keV transition is populated by the 8286-keV alpha decay from the $^{257,m}$Rf state. From the 3 counts at the expected $K$, and the 1 count obeserved at the expected $LMN+$, internal conversion peak positions in the tunnel detectors, the 750-keV transition is found to have a total conversion coefficient of $\alpha_{TOT}$=0.24(13). Using BrIcc \cite{BrIcc} the theoretical coefficients are calculated to be  0.009, 0.2, 0.035 and 0.43 for E1, M1, E2 and M2 transitions, respectively, which indicates it is mostly likely of M1 character.
The  $\alpha_K$ conversion coefficient for a 750 keV M1 and M2 transition is 0.154 and 0.315, respectively. Given the flourescence yield $\omega_K = 0.973$ \cite{TOI} and a gamma-ray detection efficiency of $\sim$30\%  at 120 keV the expected number of observed K X-rays is 5(2) or 10(3) for an M1 or an M2, respectively. Only two counts corresponding to K X-rays were observed in coincidence with the 8286-keV alpha decay thus eliminating the M2 scenario and strengthening the M1 assignment to the 750-keV transition. The weak line at 687 keV (2 counts) could be related to the 750 keV-transition and will be discussed later (NB: 687 = 750 - 63). The level at 450 keV, which deexcites via a 283-keV transition, is populated by the alpha decay of the ground state of $^{257}$Rf. As already reported in reference \cite{Streicher}, the distribution of alpha-particle energies in coincidence with this transition is large indicating that higher lying states feed into the level at 450 keV and summing effects with atomic processes occurs \cite{Theisen}. Four counts are observed in the energy range of the LMN+ conversion peak of the 283-keV transition which translates into a conversion coefficient $\alpha_{LMN+}$=0.3(2). The values calculated using BrIcc for an E1, M1 and E2 transiton are 0.01, 0.64 and 0.35, respectively. The K conversion of the 283 keV line could not be disentangled from possible contributions from unobserved low-energy transitions, however, the LMN+ conversion coefficient suggests an E2 character.

\section{Discussion}

\begin{figure}[htb]
   \centering
      \resizebox{0.95\columnwidth}{!}{%
	   \includegraphics{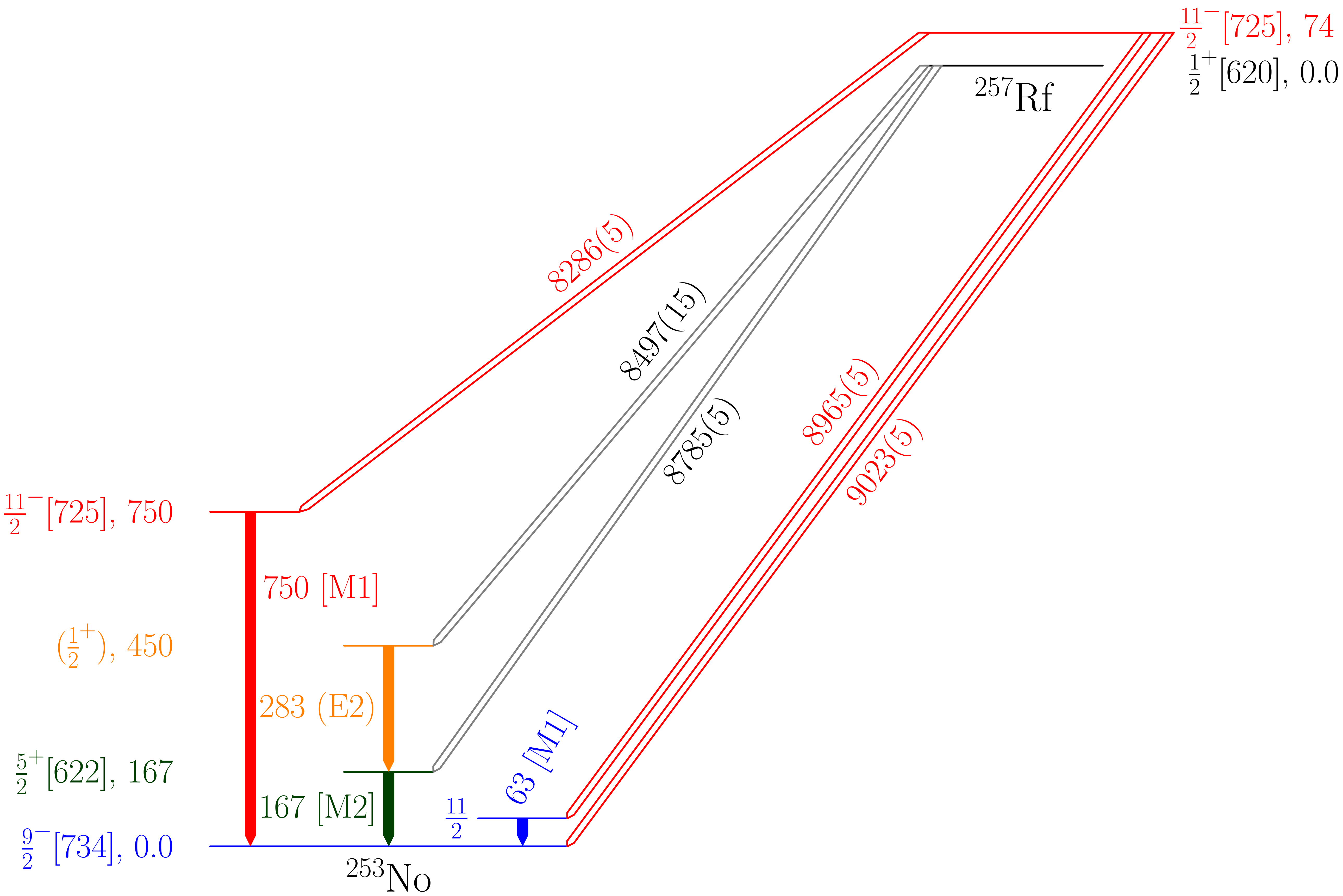}
	   }
   \caption{[Colour online] Proposed decay scheme of $^{257}$Rf. The alpha-particle energies populating the different states have a 5 keV uncertainty, except for the 8497 keV line, which is affected by summing and has an uncertainty of $\sim$ 15 keV. Bandheads are labelled by their Nilsson configurations, when known, and their exciation energies are in keV. Electromagnetic transitions are labelled with their energy in keV and the transition type is given within square brackets if known and round brackets if the assignment is in doubt.}
   \label{fig4}
\end{figure}

From Q-value considerations, the decay scheme of Fig. \ref{fig4} can be constructed. The spin of the ground state of $^{253}$No has been firmly established from a measurement of its hyperfine splitting \cite{Raeder} and has the configuration 9/2$^-$[734]. The level at 167 keV was previously assigned the configuration 5/2$^+$[622] \cite{Lopez,Streicher} - deduced from the measured  multipolarity of the transition deexciting this isomeric state. With the present data, it is not possible to determine unambiguously the spin of the state at 450 keV excitation energy. Given the available single-particle states around the Fermi level and the decay trends observed in the isotone $^{251}$Fm \cite{Asai}, the spin 1/2 has been assigned, which could correspond either to the 1/2$^+$[620] or 1/2$^+$[631] Nilsson states. Resolving the assignment would be of interest as the excitation of the 1/2$^+$[620] is related to the N=152 deformed gap as discussed in \cite{Qian}.

The M1 nature of the transition between the state at 750 keV and the ground state indicates that the state at 750 keV has a spin of either 7/2$^-$, 9/2$^-$ or 11/2$^-$. It is assigned the 11/2$^-$[725] configuration as this is the only negative parity single-neutron level near the Fermi surface other than the 9/2$^-$[734] ground state.
The excitation energy of the 11/2$^-$[725] reflects the splitting of the $j_{15/2}$ shell, from which both it and the 9/2$^-$[734] ground state emanate. In fact, both of these states are populated via $\alpha$ decay from the low lying isomer in $^{257}$Rf. Two counts are observed at 687-keV which could be a branch from the 11/2$^-$[725] to the 11/2$^-$ state of the ground state rotational band. The branching ratio expected from the Alaga rule \cite{Alaga} is $<\frac{11}{2}~1~\frac{11}{2} ~-\!1~|~\frac{11}{2}~\frac{9}{2}>^2/<\frac{11}{2} ~1~\frac{11}{2}~-\!1~|~\frac{9}{2}~\frac{9}{2}>^2$ =  0.18 compared to the observed 0.22(17). Obviously more data are required.

The $\alpha$-decay hindrance factor (HF), defined as the ratio of the experimental and theoretical partial $\alpha$-decay half-lives HF=T$_{\alpha,exp}$/T$_{\alpha,theor}$, has been shown to be a powerful spectroscopic tool when assigning intrinsic states via fine structure alpha decay \cite{Asai15}. For example, in an odd-A nucleus the favoured decay is that for which the last odd nucleon remains in the same orbital. From the $\alpha-\gamma$ coincidences, corrected for the $\gamma$-ray detection efficiency and the total number of $^{253}$No correlated $^{257}$Rf alpha decays, one obtains ratios of N$_\alpha$(8286):N$_\alpha$(8965):N$_\alpha$(9023) = 133(40):3054(789):230(60). Assuming that the alpha decay out of the $^{257,m}$Rf state proceeds via only these decays with a branching ratio of 0.65 (0.35 proceeds to the ground state via internal decays) a lower limit of the alpha hindrance factor of 2.1(7), 13(4) and  241(69) have been calculated using the code ALPHAD-v2d \cite{ALPHAD}, with $r_0$=1.468 fm, for the 8286-, 8965- and 9023-keV transitions respectively. The favoured alpha decay firmly establishes the configuration of the $^{257,m}$Rf state as the 11/2$^-$[725].


From N=153 systematics, and the $\alpha$ decay pattern of $^{261}$Sg \cite{Antalic}, the ground state in $^{257}$Rf has been assigned the 1/2$^+$[620] configuration. In the isotone $^{251}$Cf the excitation energies of the 3/2$^+$, 5/2$^+$ and 7/2$^+$ members of the ground state rotational band have been firmly established to be 24.8, 47.8 and 105.7 keV respectively \cite{Ahmad}.
Assuming the ground state rotational band to be similar in $^{257}$Rf  would result in the 5/2$^+$ member of a rotational band lying approximately 30 keV below the alpha decaying isomer while the 7/2$^+$ state would be above the isomer.
The internal decay from the 11/2$^-$[725] band-head would therefore proceed via an E3 transition of approximately 30 keV to the 5/2$^+$.
Using a branching ratio, B, of 35$\%$  as suggested by the fit to the half-life of the ground state discussed above, the partial $\gamma$ half-life of the 11/2$^-$ state is $t_{1/2}^{\gamma}\!=\!t_{1/2}.(1+ICC)/B\sim7.21\times10^6$s, where ICC is the total internal conversion coefficient for a 30 keV E3 transition. The Weisskopf estimate for such a transition is  $t_{1/2}^{W}\!=\!1.41\times10^4$s indicating that the transition is retarded by a factor $F^W = t^\gamma_{1/2}/t^W_{1/2}\sim 510$ and has a reduced hindrance factor of $f_\nu = (F^W)^{1/\nu}\sim$23, where  $\nu=\Delta$K-$\lambda$, $\Delta$K is the difference in the K quantum number between the initial and final states and $\lambda$ is the transition multipolarity. Recently Kondev et al. \cite{kondev} have re-evaluated the L\"obner systematics \cite{Lobner}, and while they mention the limited data for E3 decays, using their fit to known data results in an expected $f_\nu(E3) \sim 79$. Given the approximations made concerning the E3 transition energy, the uncertainty in the ICC so close to  L2-shell binding energy and the errors in the fit in \cite{kondev}, the agreement is reasonable and indicates that the meta-stable state at 74 keV in $^{257}$Rf is in fact a low lying K-isomer.

Interestingly, a weak 750 keV transition has been observed in the decay of the 674$~\mu$s high-K isomer in $^{253}$No \cite{Hessberger3,Lopez2,Antalic}. Since some of the high-energy lines observed in the decay of this isomer have been seen at the target position in a prompt spectroscopy experiment \cite{Eeckhaudt}, the decay of the isomer must go through an intermediate structure, which emits promptly the aforementioned transitions. An intermediate structure based on the coupling of the odd 9/2$^-$ neutron to a 2-quasiproton 3$^+$ excitation was put forward in reference \cite{Lopez2}. Given the present data, a different scenario can be constructed. By building a rotational band on top of the 11/2$^-$ state of Fig. \ref{fig4}, with a similar sequence as the one reported by Rissanen et al. in $^{257}$Rf \cite{Rissanen}, most of the high-energy gamma lines seen in the decay of the isomer (at 614, 703, 714, 779 and 802 keV) can be reinterpreted as decays from members of the 11/2$^-$ band to states of the ground state band. This scenario is in part supported by the fact that the 11/2$^-$ state is predicted to carry a significant octupole phonon content \cite{Shirikova}, which could explain why the decay occurs all through the band.

\section{Conclusion}

The properties of the alpha decay of $^{257}$Rf have been re-investigated with the GABRIELA detection setup. The half-lives extracted for the two known alpha-emitting states have been measured. The half-life of the low-lying isomer is in agreement with previously-published values, while the that of the ground state is found to be significantly higher. This fact can be attributed to feeding of the ground state from the isomer with a branching ratio of 35(5)$\%$. The alpha decay has revealed a significant branch, with a hindrance factor of 2, from the low lying isomer in $^{257}$Rf to a new state in $^{253}$No, which decays by an M1 transition to the 9/2$^-$ ground state. This decay path has the same Q-value as other branches observed to the first excited state of the ground state rotational band and the ground state itself, giving confidence in the established decay scheme. Building upwards from the known configuration of the ground state in $^{253}$No we can assign the 11/2$^-$[725] configuration to both the excited state at 750-keV  in $^{253}$No and the alpha emitting meta-stable state in $^{257}$Rf. The 11/2$^-$[725] state in N=151 isotones decreases in excitation energy as Z increases, as a fine structure alpha decay study places it at $\sim$ 600 keV in $^{255}$Rf \cite{Antalic2}. A similar trend is observed in N=153 isotones, where the 11/2$^-$ state steadily decreases in excitation energy from $^{251}$Cf onwards  to become the ground state in $^{259}$Sg. This behaviour is explained by theoretical calculations using the Nilsson-Strutinsky approach based on a Woods-Saxon model \cite{Cwiok} and arises because of changes in deformation.

The possible implication of the 11/2$^-$ state in decay of the known high-K isomer in $^{253}$No has been discussed. This scenario, however, needs verification with a dedicated investigation as the complete decay scheme from the isomer needs to be established.

\section{Acknowledgements}

This work was supported by the French National Research Agency (Project Nos. ANR-06-BLAN-0034-01 and ANR-12-BS05-0013), the IN2P3-JINR collaboration Agreement No. 04-63, the Russian Foundation for Basic Research  (RFBR), Contracts No. 08-02-00116, 17-02-00867 and 18-52-15004, and by the Slovak Research and Development Agency (contract number APVV-18-0268). The authors thank the U400 cyclotron and ion-source operations staff for providing stable high intensity beams.

%

%
%

\end{document}